\documentclass[conference]{IEEEtran}
\IEEEoverridecommandlockouts
\usepackage{cite}
\usepackage{amsmath,amssymb,amsfonts}
\usepackage{algorithmic}
\usepackage{graphicx}
\usepackage{textcomp}
\usepackage{xcolor}

\begin{document}
\title{CausalML: Python Package for Causal Machine Learning \thanks{* Authors listed alphabetically. All authors are from Uber Technologies Inc.}}

\author{Huigang Chen*, Totte Harinen*, Jeong-Yoon Lee*, Mike Yung*, Zhenyu Zhao*}

\maketitle

\begin{abstract}
\textit{CausalML} is a Python implementation of algorithms related to causal inference and machine learning. Algorithms combining causal inference and machine learning have been a trending topic in recent years. This package tries to bridge the gap between theoretical work on methodology and practical applications by making a collection of methods in this field available in Python. This paper introduces the key concepts, scope, and use cases of this package. 
\end{abstract}


\maketitle

\section{Introduction}
\textit{CausalML} is a Python package that provides a suite of uplift modeling and causal inference methods using machine learning algorithms based on cutting edge research. The traditional causal analysis methods, such as performing t-test on randomized experiments (a.k.a. A/B testing) can estimate the Average Treatment Effect (ATE) of the treatment or intervention. However, in many applications, it is often desired and useful to estimate these effects at a more granular scale. \textit{CausalML} enables the end user to estimate the Conditional Average Treatment Effect (CATE), which is essentially the effect at the individual or segment level. Such estimates can unlock a large range of applications for personalization and optimization by applying different treatment to different users. 

One key modeling technique enabled by \textit{CausalML} is uplift modeling. Uplift modeling \cite{Grimmer2017-rl, Guelman2015-qe, Gutierrez2016-co, Kunzel2017-ko, Rzepakowski2012-br, Soltys2015-be, Wager2015-sd, Zaniewicz2013-rt, Zhao2017-kg, nie2017quasi} is a causal learning approach to estimate the individual treatment effect for an experiment.  This allows the end user to measure the incremental impact of a treatment (such as a direct marketing action) on an individual's behaviour using experimental data. For instance, if a company is choosing between multiple product lines to up-sell / cross-sell to its customers, it can utilize \textit{CausalML} as a recommendation engine to identify products that achieve the greatest expected lift for any given user.

It is worth mentioning that this package is not designed to replace the standard randomized experiment approach for drawing causal inference. In many scenarios, it is essential to carry out randomized experiments to evaluate the ATE for business decisions. While uplift modeling can be applied to both experimental data and observational data, the current implementation of the uplift modeling is encouraged to be applied to the data from the randomized experiment. Applications to observational data where the treatment is not assigned randomly should take extra caution. In non-randomized experiment, there is often a selection bias in the treatment assignment (a.k.a. confounding effect). One main challenge is that omitting potential confounding variables in the model can produce biased estimation for the treatment effect. On the other hand, properly randomized experiments do not suffer from such selection bias, that provides a better basis for uplift modeling to estimate the CATE (or individual level lift). 

There are a few related  packages. In R, the uplift, grf, and rlearner packages implement the Uplift Random Forest, Generalized Random Forest, and R-learner methods respectively. In Python, the package DoWhy is focused on structuring the causal inference problem through graphical models  based on Judea Pearl's do-calculus and the potential outcomes framework. The recently released EconML Python package implements heterogeneous treatment effect estimators from econometrics (such as instrumental variables) and machine learning methods . Another Python package Pylift implements one meta-learner for uplift modeling. The contribution of the current implementation of \textit{CausalML} package is providing an one-stop-shop for uplift modeling techniques (8 models so far) with a set of support functions in Python. For example, according to our knowledge, the Uplift Random Forest methods and R-learner are made available in an open source Python package for the first time. In addition, we have implemented innovative methods developed in house, such as meta-learners for multiple treatment groups optimization.

\section{Why CausalML}
In recent years, the intersection of causal inference and machine learning has become an active area of research. Based on our experience at Uber, we believe there are going to be a number of important practical applications emerging from this research. With the \textit{CausalML} package, we aim to make these applications accessible to a wider audience. Our ultimate goal is to build a one-stop shop for machine learning for causal inference.

The first area of focus for the package is the area of uplift modeling. We believe it is a methodology that can serve an important purpose in business, science and elsewhere. With the first version of the \textit{CausalML} package, our goal is to democratise the uplift modelling methods that are currently available in only academic papers or in disparate statistical packages.

To do so, we offer a consistent API that makes running an uplift algorithm as easy as fitting a standard classification or regression model. Model performance can be evaluated using the included metrics and visualisation functions such as uplift curves. The first version of \textit{CausalML} implements eight state of the art algorithms for uplift modelling (see Figure. 1).

\begin{figure}
\centering
\includegraphics[width=0.49\textwidth]{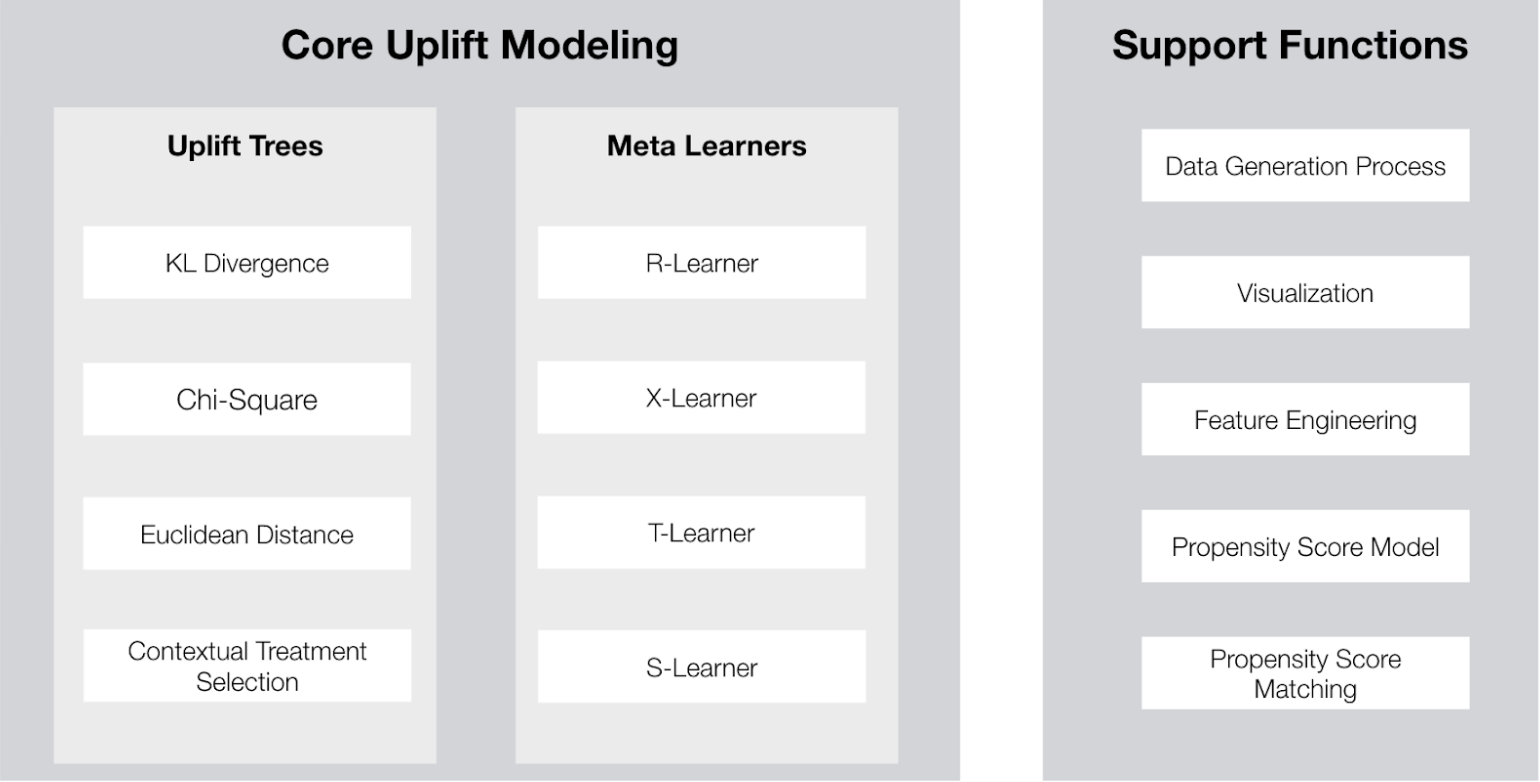}
\caption{\textit{CausalML} package diagram}
\label{fig:package_diagram}
\end{figure}

Further, we have built the package flexible in terms of the types of outcome variables that can be modelled, covering both regression and classification type tasks. The package also contains algorithms that can be used with data from experiments with multiple treatment groups. Our hope is that these features of flexibility, generality and ease of use will make \textit{CausalML} the tool of choice for uplift modellers in the future.

\section{What Problems Can CausalML Solve?}
\textit{CausalML}’s use cases include, but are not limited to, targeting optimization, engagement personalization and causal impact analysis.

\subsection{Targeting Optimization}
We can use \textit{CausalML} to target promotions to those with the biggest incrementality. For example, at a cross-sell marketing campaign for existing customers, we can deliver promotions to the customers who would be more likely to use a new product specifically given the exposure to the promotions, while saving inboxes for others. An internal analysis showed that targeting only 30\% of users with uplift modeling could achieve the same increase in conversion for a new product as offering the promotion to all customers.

\subsection{Causal Impact Analysis}
We can also use \textit{CausalML} to analyze the causal impact of a particular event from experimental or observational data, incorporating rich features. For example, we can analyze how a customer’s cross-sell event affects long term spending on the platform. In this case, it is impractical to set up a randomized test because we do not want to exclude any customers from being able to convert to the new product. Utilizing \textit{CausalML}, we can run various ML-based causal inference algorithms, and estimate the impact of the cross-sell on the platform.

\subsection{Personalization}
\textit{CausalML} can be used to personalize engagement. There are multiple options for a company to interact with its customers, such as different product choices in up-sell or messaging channels for communications. One can use \textit{CausalML} to estimate the effect of each combination for each customer, and provide optimal personalized offers to customers.

\section{Future Development}
\textit{CausalML} is actively maintained and developed by the Uber development team \footnote{The Uber development team for \textit{CausalML} includes Huigang Chen, Totte Harinen, Jeong-Yoon Lee, Mike Yung, and Zhenyu Zhao (alphabetically).} for \textit{CausalML}. One area of development is to further improve the computational efficiency for existing algorithms in the package.  In addition, we plan to add more state-of-the-art uplift models as needed in the future. Besides uplift modeling, we are also exploring more modeling techniques in the intersection of machine learning and causal inference, with a goal of solving optimization problems. 

We welcome everyone to try out \textit{CausalML} on different use cases, and invite you to share your feedback with us. If you are interested in contributing to the development of this package, please read our code of conduct and follow contributing guidelines. 

\section{Acknowledgement}
We would like to extend our appreciation to our colleagues at Uber who have contributed to and supported this open source project, including Mert Bay, Fran Bell, Natalie Diao, Shuyang Du, Neha Gupta, Candice Hogan, Yiming Hu, James Lee, Paul Lo, Yuchen Luo, Lance Mack, Vishal Morde, Jing Pan, Hugh Williams, Yunhan Xu, and Yumin Zhang. 
We would also like to thank external contributors for this project: Peter Foley, Florian Wilhelm,  Steve Yang, and Tomasz Zamacinski.

\bibliographystyle{./IEEEtran}
\bibliography{./uplift_references}
\end{document}